# Marginal energy intensity of water supply

Yang Liu[1], Meagan S. Mauter[1,2]*

[1]Department of Civil and Environmental Engineering, Stanford University, Stanford, CA 94305, USA.

[2]National Alliance for Water Innovation, Lawrence Berkeley National Laboratory, Berkeley, CA 94720, USA.

*Corresponding author. E-mail: mauter@stanford.edu

**Abstract:** Reducing global carbon emissions will require diverse industrial sectors to use energy more efficiently, electrify, and operate intermittently. The water sector is a transformation target, but we lack energy quantification tools to guide operational, infrastructure, and policy interventions in complex water sourcing, treatment, and distribution networks. The marginal energy intensity (MEI) of water supply quantifies the location-specific, instantaneous embedded energy in water delivered to consumers. We describe the first MEI algorithm and elucidate the sensitivity of MEI to generalizable water system features. When incorporated in multi-objective operational and planning models, MEI will dramatically increase the energy co-benefits of water efficiency, conservation, and retrofit programs; maximize energy flexibility services that water systems can deliver to the grid; and facilitate full cost recovery in distribution system operation.

## Introduction

Energy consumption by the water sector presents both challenges and opportunities in efforts to mitigate and adapt to climate change. Water and wastewater utilities account for up to 6% of regional electricity consumption (*1, 2*), with water-stressed or water-impaired regions increasingly dependent on energy intensive desalination processes to secure their water supply. On the other hand, water systems are ideal providers of energy flexibility services to the grid because they are energy intensive, centrally managed, and robust to variable operation (i.e., storage and redundancy are inherent features of these systems) (*3, 4*). Optimizing water grid system operation, prioritizing infrastructure investments, and setting policies to minimize energy consumption and maximize energy flexibility services requires a detailed understanding of the energy intensity of these water supply systems (WSS).

A large body of past work has quantified the average energy intensity of water supply at the system-level (*2, 5–10*). In general, these studies sum the daily or yearly energy demand of system components and divide by the total volume of water supplied. While more recent work acknowledges the wide variation in energy intensity across different pressure zones (*11*) and locations (*12*), critical limitations inherent in past modeling approaches have prevented accurate accounting of the spatial or temporal variability in energy intensity (Supplementary Note 1). The range of approaches for calculating spatially or temporally averaged energy intensity are useful for attributing energy consumption across the water sector and comparing it to that of other sectors, but these inventory techniques must not be used to optimize instantaneous system operation or inform long-term energy policy or infrastructure upgrades in the water sector.

Instead, WSS regulators, operators, and consumers need consequential accounting techniques that provide location-specific, instantaneous insight into WSS energy intensity. We propose a new metric, the marginal energy intensity (MEI) of water supply, representing the energy intensity of the next unit of water consumed at a specific location and time. MEI captures the high spatial and temporal variability inherent in WSS energy intensity and is essential for guiding energy efficient WSS expansion, retrofit, repair; maximizing the effectiveness of policies and consumer-based incentive programs; and determining the magnitude of energy flexibility services that WSS might provide to the electricity grid. MEI will offer further value as WSS evolve toward a hybridized centralized/decentralized model where water users meet demand through a combination of traditional WSS networks and localized reuse (*13*, *14*).

This paper presents a computationally rigorous approach for computing the MEI of WSS. We begin by detailing the steady-state and dynamic energy attributes of complex WSS from transmission (i.e., conveyance) through treatment and distribution. We then describe a disaggregated computational framework for computing the location- and time-dependent energy intensity of water and apply this framework to a case study with three sources and 330 water consumer nodes. We evaluate the variability of MEI as a function of location and time of day, as well as the relative magnitude of transmission, treatment, and distribution contributions to MEI across the base case WSS. Finally, we perform a sensitivity analysis of base case assumptions to assist in generalizing insights from the base case model to a diverse array of WSS. The proposed computational framework represents the first true source(s)-to-consumer marginal energy intensity of water supply in a generic multi-source WSS with substantial water storage capacity and represents a transformative tool for optimizing WSS operation and guiding climate change adaptation and mitigation policies in the water sector.

## Marginal energy intensity of water supply

The marginal energy intensity of a unit of water consumed by an end user at a specific time is computed by determining the instantaneous energy consumption of WSS components (e.g., pumps) and apportioning the fractional use of those components to individual consumers via a flow backtracking method. A conventional WSS will consume energy in water transmission, treatment, and distribution (Fig. 1). Since water transmission is often synchronized with near steady-state treatment operations, this version of our model approximates the energy consumption of both water transmission and treatment as steady-state. Where treated water is injected into the water distribution network (WDN), this steady-state model intersects with a dynamic model for water distribution.

The electricity consumption of water distribution systems is highly variant as a result of non-steady state electricity prices, non-linear pump curves, and complex relationships between water demand, water path, and frictional losses in pipes. We implement a genetic algorithm (see Methods section, Supplementary Note 2, and Fig. S1) with hourly resolution to mimic the behavior of rational WSS operators that schedule water pumping operations to minimize their operating costs (*15–20*). We then use flow rates, pressures, and mechanical efficiencies simulated from the pumping schedule to calculate the energy consumption of each active device (e.g., pumps, treatment processes) or energy dissipation by each passive component (e.g., pipes, valves) along the flow paths between each source-consumer pair.

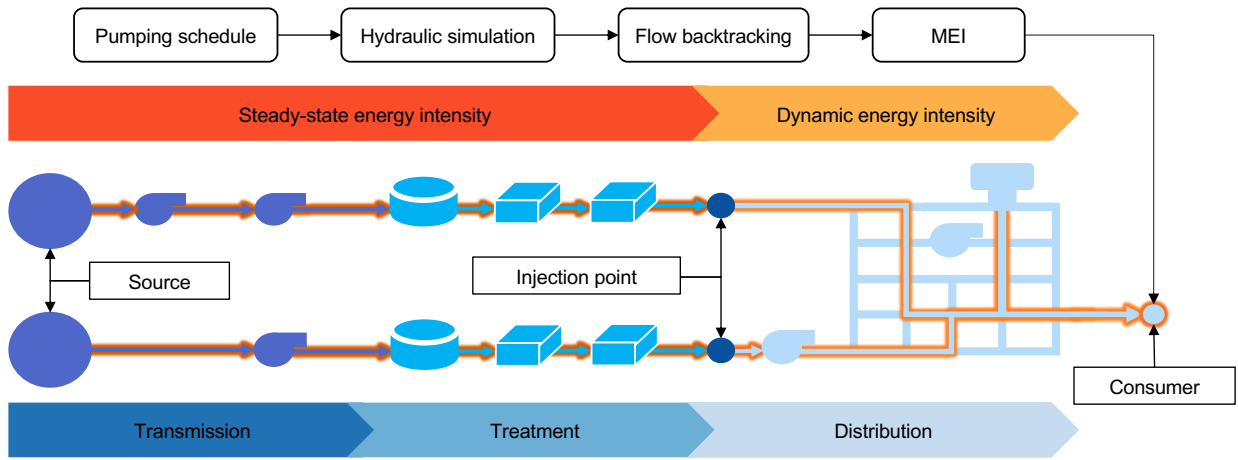

**Fig. 1. Method for computing the marginal energy intensity of water supply.** MEI accounts for energy consumed in water transmission, treatment, and distribution. This study assumes transmission and treatment to be steady-state, but alternative formulations of this model can be applied to systems with highly time-variant energy intensities for transmission or treatment. The location- and time-specific MEI value for a given consumer can be calculated by first backtracking the delivered water to its injection points in the WDN and then backtracking from the injection points to the raw water sources. The backtracking requires a simulation of water flows in the WDN under a feasible pumping schedule.

Next, we decompose the water received by each consumer by source using a flow backtracking algorithm that calculates the volume of water received from each source by each consumer during each time step. This algorithm assumes that each node in the WDN behaves as a perfect mixer of upstream water and allows tanks (i.e., water storage) to serve as the primary source of water supply during hours when pump load is low. To accurately account for the energy consumed to fill up tanks in previous time periods, we calculate the time averaged pre-injection energy intensity of each tank as a constant whose value depends on the origin of the stored water.

Finally, we compute the MEI of each consumer by summing the time-dependent energy intensities of the devices and components along the time-specific flow path of the delivered water. Marginal energy intensity values are computed at the node and hourly levels and reported at the node and either the daily or the hourly level in this work, but there is no limitation on the spatial or temporal resolution of MEI calculations using this method.

## Application to a multi-source WSS

We apply our computational framework for calculating MEI values to a WDN model (Fig. 2A) adapted from a functioning system in Kentucky, U.S (*21*, *22*). The un-skeletonized WDN has 87.7 km of pipes and serves 2682 consumers with approximately 1.04 million gallons (4581) of water per day, while our skeletonized version aggregates this demand to 330 consumer nodes. Additional WDN specifications are provided in Supplementary Note 3 (Table S1, Fig. S2-S3). This WDN has three injection points, I1, I2, and I3. The pre-injection energy intensities of these sources represent the average energy intensity of a local groundwater source, a local surface water source, and a distant source requiring energy intensive inter-basin water transfer, respectively (*23*).

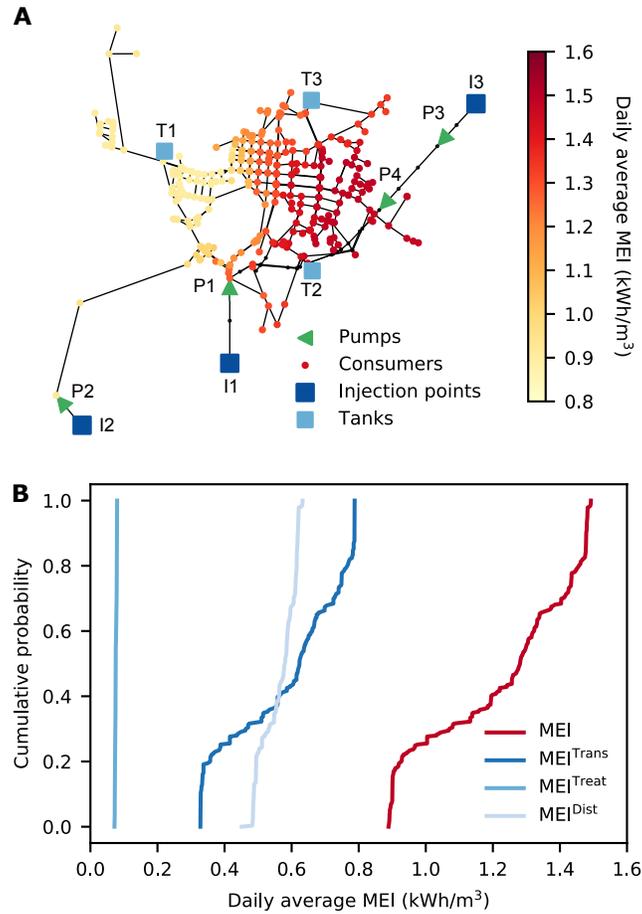

**Fig. 2 Daily average marginal energy intensity of the base case water distribution network.
A.** The base case WDN serves 330 consumer nodes with four pumps (P1-P4), three injection points (I1-I3), and three tanks (T1-T3). The pre-injection energy intensities of the three injection points are 0.4 kWh/m$^3$, 0.11 kWh/m$^3$, and 1.05 kWh/m$^3$, respectively. The length of the long pipes immediately downstream I3 are shortened in this figure for ease of visualization (Fig. S2). **B**. Cumulative distribution functions representing the daily average MEI for the base case WDN and the relative contributions of transmission, treatment, and distribution of water to this total MEI value. Steeper slopes represent less variation in the daily average MEI across individual consumer nodes.

We compute the base case MEI for a defined node-level water consumption profile, time-of-use electricity price profile, pipe friction factors, and specified target fraction of water sources (Fig. 3). In the base case scenario, the target fraction of daily injection through I1, I2 and I3 is 25%, 25% and 50%, respectively. In practice, source water fractions are determined through long-range planning processes that account for the availability and cost of each source.

Under electricity price pattern EP0 (Fig 3A), the pumping activities are optimally scheduled to take place during 10:00-18:00 and 23:00-5:00. From 11:00 to 14:00, both injection points I2 and I3 are directly contributing to the water supply. Similarly, injection through I1 and I2 are synchronized from 3:00 to 4:00. Besides these two periods, the pumping of water through each injection point is desynchronized from those through other injection points. In this three-source WDN, periods of synchronization produce a much wider range of instantaneous MEI values (Fig. 3C) because consumers are receiving water from injection points with vastly different pre-injection energy intensities. When electricity prices are high, pumps are turned off and water is supplied solely from tanks. During these 'no-pumping' periods (6:00-10:00, 18:00-23:00 and 5:00-6:00), the node-specific MEI values are clustered around three values that reflect the pre-injection energy intensities of the three tanks in the system (T1, T2, and T3). Since the tanks mix water from all three sources, consumers who receive more water from tanks tend to have less extreme MEI values than those who receive more water directly from a high- or low-energy intensity source. In general, the magnitude of spread in MEI values indicates the similarity in source-to-consumer node flow paths among the consumer nodes. Multiplying hourly MEI values by hourly water consumption provides the daily average MEI values for the 330 consumer nodes (Fig. 2A), which range from 0.89 kWh/m$^3$ to 1.49 kWh/m$^3$. The system average daily MEI value, which equals the system average energy intensity of water supply, is 1.23 kWh/m$^3$. Energy consumed in water distribution accounts for 46% of this average energy intensity, or 0.56 kWh/m$^3$, with 0.19 kWh/m$^3$ consumed by energy dissipation during the final reaches of the distribution system. The high pre-injection energy intensity of I3 leads to higher daily average MEI values for consumer nodes receiving a higher fraction of water from I3 (Fig. 2A). A more wholistic view of the relative contribution of transmission, treatment, and distribution to the range of MEI values for the base case scenario is provided in Fig. 2B. Transmission- and distribution-associated components are the primary determinants of total MEI for this base case scenario, but this trend would not necessarily hold for systems with energy intensive unit processes, such as reverse osmosis (*24, 25*).

## Sensitivity analysis

The instantaneous MEI value of a single consumer node is a marginal and, thus, insensitive to modest perturbations in the instantaneous water consumption profile (Fig. S6 and Supplementary Note 6). This marginality makes MEI valuable for assessing incremental changes in water consumption behavior associated with operational, infrastructure, or policy changes. Therefore, to elucidate factors influencing MEI, we vary system-wide water consumption, electricity price, pipe roughness, and the daily injection fraction from each source and then compare the resulting distributions of daily average MEI values across the WDN.

Throughout the sensitivity analysis, the pumping load profile (Fig. 3D) and source-specific daily water injection volumes (Fig. 3E) are calculated for each scenario from the optimized pumping schedule. According to Fig. 3F, daily average MEI values for the 330 consumer nodes are relatively insensitive to changes in water demand (Fig. 3F, scenarios D$^-$, D$^+$) and the temporal shift of pumping load due to changes in the electricity price pattern (Fig. 3F, scenarios EP1 and EP2). These scenarios have little influence on the energy intensity of the flow paths from the injection points to the consumers, but instead influence the duration and timing of the periods in which each source-to-consumer path is activated to transport water. In contrast, daily average MEI values are highly sensitive to pipe roughness, which can also serve as a proxy for distribution network size (discussed in Supplementary Note 7). The daily average MEI for scenario R$^+$, with a 50% increase in roughness leads to an average of 8.7% (0.11 kWh/m$^3$) increase in daily average MEI among the 330 consumer nodes. This increase is attributed both to head losses due to pipe roughness and the decreased mechanical efficiency of pumps operating at higher heads and lower discharge rates. Similarly, the curve of daily average MEI values shifts leftward (i.e., decrease) by 13.0% (0.16

kWh/m$^3$) when the pipe roughness decreases by 50%. Daily average MEI values are also highly sensitive to the target fraction of daily water supply from each source. As daily injection through I3 decreases or increases to 30% or 70% in scenarios I3$^-$ and I3$^+$, respectively, the daily average MEI values shift by -20.7% and +16.9% on average. The rightward (i.e., +16.9%) shift reflects both the higher pre-injection energy intensity of I3 and the energy intensity of pumps P3 and P4 that link I3 to the consumer nodes.

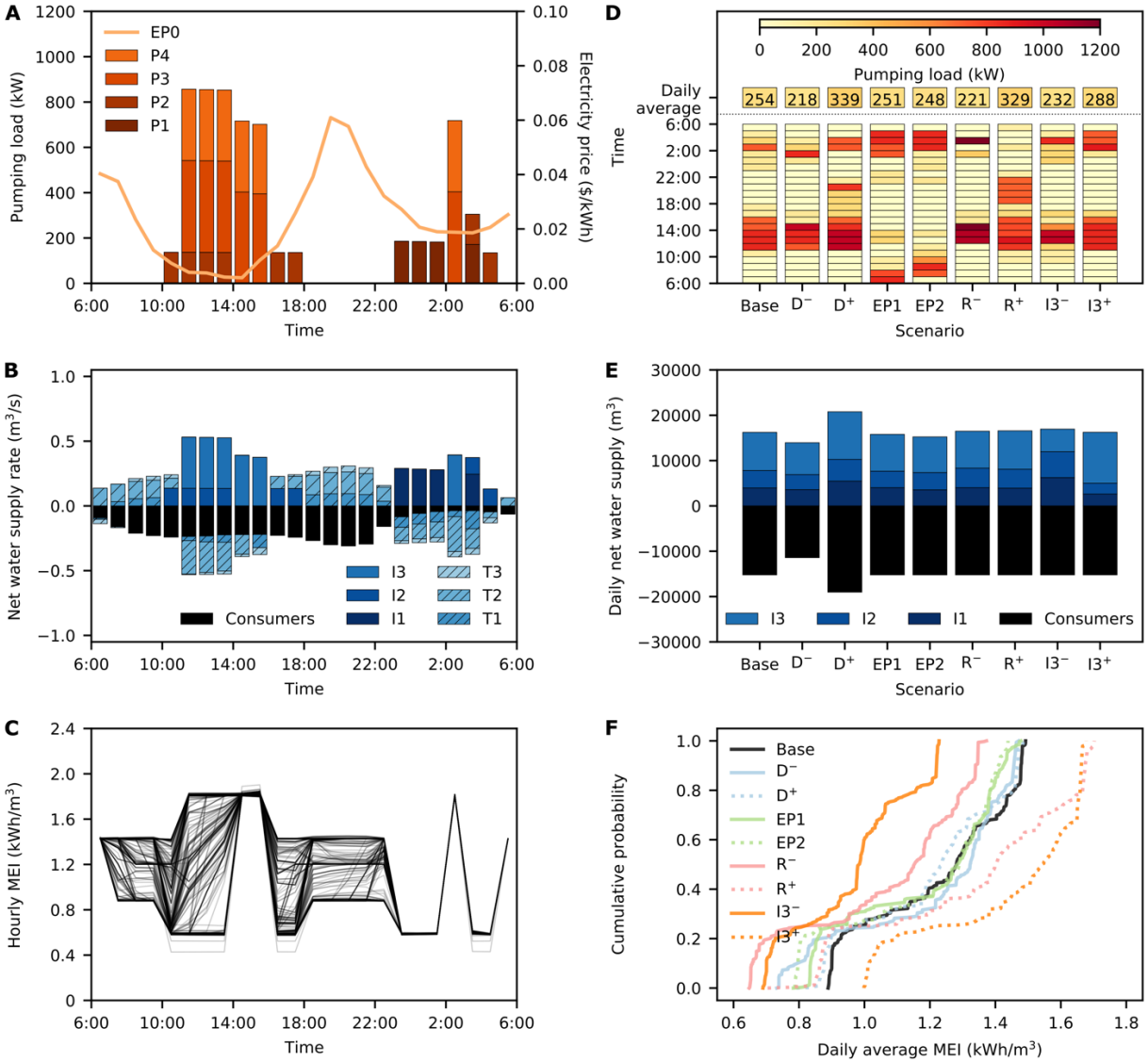

**Fig. 3. Hourly base case results and sensitivity analysis.** Temporally resolved results from the base case scenario (a-c) and sensitivity analysis for eight additional scenarios (d-f). 'Base' is the base case scenario. 'D$^-$' and 'D$^+$' are scenarios where the hourly water consumption rates (i.e., demand) of all consumer nodes are decreased or increased by 25%, respectively. 'EP1' and 'EP2' are scenarios where alternative electricity price profiles (Fig. S4 in Supplementary Note 4) are used to optimize the pumping schedule. 'R$^-$' and 'R$^+$' are scenarios where the roughness of all pipes is decreased or increased by 50%, respectively. 'I3$^-$' and 'I3$^+$' are scenarios where the target percentage of daily water injection through I3 is adjusted from 50% in the base case to 30% or 70%, respectively. **A**. Hourly load profile of the optimal pumping schedule that minimizes the electricity cost under electricity price profile EP0. **B**. Hourly water injection through each injection point and the net water flows into tanks and consumer nodes. **C**. Hourly MEI of consumers. **D**. Hourly load profiles of the optimal pumping schedules in nine scenarios. **E**. Daily water injection through each injection point and daily water demand of consumers nodes in nine scenarios. **F**. Distribution of daily average MEI of consumer nodes in nine scenarios. The hourly MEI values of all scenarios are displayed in Fig. S5 (Supplementary Note 5).

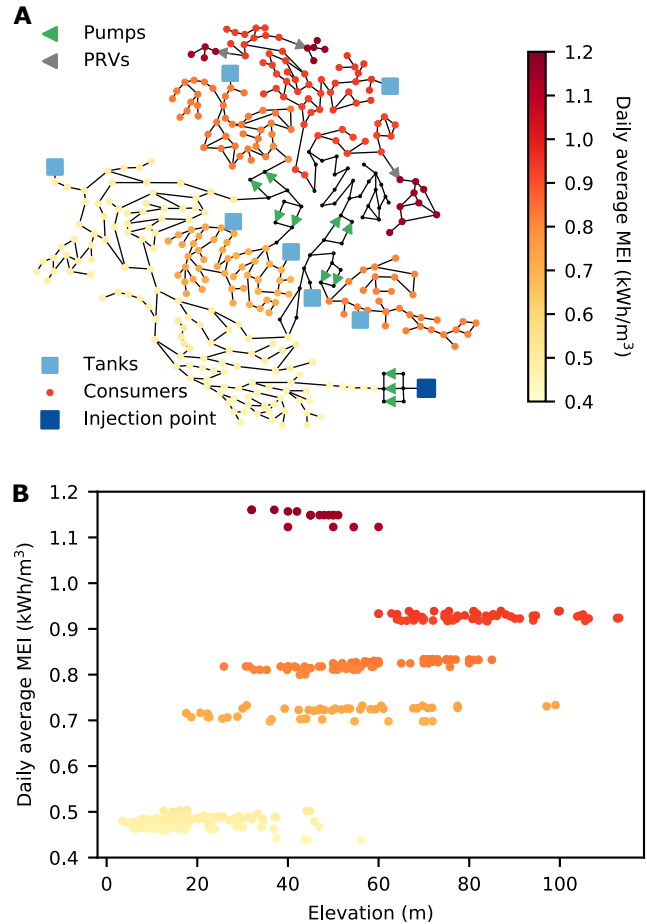

**Fig. 4. Relationship between MEI and consumer node elevation. a**. Layout of a WDN with 11 pumps (three primary pump stations, eight booster pumps), three pressure reducing valves (PRVs), 334 consumers, one injection point, and seven tanks. **b**. Daily average MEI values of consumer nodes located at different elevations. The daily average MEI legend applies to both panels.

Finally, besides the factors captured by the eight scenarios shown in Fig. 3, the daily average MEI value of a consumer node is highly correlated to its pressure zone. WSS serving regions with significant topographic variability are segmented into pressure zones. For example, the East Bay Municipal Utility District serving Alameda County in California operates their WSS using approximately 130 different pressure zones to ensure adequate water pressure for consumers. Most zones are separated from the rest of the network by devices allowing only single-direction flow (e.g., booster pump, check valve) and contain a dedicated water tank. To isolate the relationship between the elevation of a consumer node and its daily average MEI, we apply the same computational framework shown in Fig. 1 to a representative single-source WSS (Fig. 4A) with a pre-injection energy intensity of 0.3 kWh/m$^3$. This WDN is adapted from the 'c-town network' (26) and additional details are provided in Supplementary Note 8. Modifications to the WDN are listed in Table S2. All simulations for this WDN use the pumping schedule (Fig. S7) optimized with electricity price pattern EP0.

The daily average MEI values of the 334 consumer nodes in this single-source WDN are clustered into five groups (Fig. 4B) that capture the location of consumer nodes relative to booster pumps (green). The operating head and energy intensity of the booster pumps are typically established by the highest elevation consumer node within the pressure zone, such that supplying low elevation nodes within this pressure zone still results in high daily average MEI values. If the resulting elevation head exceeds the pressure tolerance of pipes at low elevations, pressure reducing valves (PRVs, represented in grey) dissipate the excess energy. In this WDN, the highest daily average MEI values are associated with these distant, dead-end, low-elevation consumer nodes downstream of high-elevation ones. As such, we note that consumer elevation alone is a poor predictor of MEI values. Furthermore, while higher-elevation consumers are served by higher pressure zones that are correlated with high MEI values in this single-source WDN, this relationship does not hold for distribution systems with multiple sources, with injection points located at different elevations, or for systems involving treatment processes of varying energy intensities. A flow backtracking algorithm is essential for robust calculation of the MEI values in a distribution system.

## Discussion

Water systems are facing unprecedented environmental, financial, and regulatory pressure. Climate change and growing urban populations are limiting the availability of clean, fresh water sources and necessitating that water users either transport water long distances or tap nontraditional water supplies that require energy intensive treatment. Aging infrastructure systems demand significant capital investments in maintenance and retrofit, but many WSS systems are unable to recover the full costs of system operation and asset management. Finally, regulators are working to safeguard human heath by tightening water quality standards, while also capping the carbon emissions of the water sector. In short, there is a critical need for optimization tools that balance these competing pressures and aid water supply systems in strategically preparing for the future. The energy intensity of water supply is central to each of these pressures, but we are unaware of prior work quantifying spatially and temporally resolved MEI critical to optimizing water system management. This work introduces the concept of MEI, comprehensively details a flow backtracking method for computing MEI, and applies this method to two distinctly different water systems.

Intuitively, MEI increases when WSS rely on energy intensive sources and treatment trains, when there is significant friction loss in distribution networks, and when a consumer is located in the same pressure zone as high elevation consumers. Less intuitively, consumer MEI fluctuates quite dramatically over the course of the day and the timing of fluctuation is a strong function of time-of-use electricity prices, water storage volumes of the tanks, and water demand. The resulting complexity of these interacting factors suggests a continuous role for real-time WSS management optimization platforms that minimize the cost and energy intensity of water supply.

WSS management optimization platforms that calculate MEI could also be used to guide operational, planning, and policy decisions. As an asset management tool, MEI could aid in the design of incentive programs that temporally shift flexible consumer water use (e.g., by irrigation systems) or accelerate the replacement of water-intensive appliances (e.g., high flush toilets) in locations with high MEI. As an asset maintenance tool, MEI could aid utilities in prioritizing replacement of leaking pipes or identify areas where rainwater capture or distributed reuse might substantially reduce the energy intensity of water supply. As a financial tool, MEI and the underlying flow backtracking tool could aid utilities in full cost recovery of water distribution system maintenance and pumping costs.

Finally, MEI will play a critical role in identifying components and users capable of providing demand response capacity to the electricity grid. Past work has identified WSS as ideal participants for DR provision due to their large pumping load and large water storage capacity (*3*, *4*, *27*), but such work focused exclusively on decoupling water supply and demand through tank storage. The MEI calculations performed in this work suggest that significant additional DR capacity exists if WSS systems can leverage temporal flexibility in water demand. Temporally shifting the water consumption of high-MEI consumers allows energy intensive WDN components to shift their load without jeopardizing the continuity of water supply.

While MEI is one critical metric in WSS operation, it is not the only one. Cost, reliability, water quality, carbon intensity and other metrics determine optimal WSS design and operation. Fortunately, the flow backtracking method detailed in this paper enables facile determination of these other factors. Indeed, calculating these spatial and temporally resolved metrics and incorporating them into multi-objective optimization models for water system management is a crucial next step in WSS research.

## Methods
### Pumping schedule
Any feasible pumping schedule is sufficient for computing MEI using the proposed computational framework (Fig. 1), but here we implement a heuristic optimization method (i.e., genetic algorithm) to solve for near-optimal pumping schedules that minimize the electricity costs. We include three penalty functions to reflect other common factors in determining the pumping schedule of a water distribution network (WDN). First, the reliability of water supply is ensured by penalizing solutions where the final water level of a tank is lower than its initial level. Second, we penalize solutions that deliver water at pressures lower than 20 psi. Finally, we penalize large deviation (i.e., >2%) from the planned percentage of water injection through each injection point to ensure long-term reliability of water supply.

Following past genetic algorithms (*15–20*), we outsource the handling of hydraulic constraints in solving the pump scheduling problem to WNTR (*28–30*), the Python version of the widely used EPANET software (*31*). The genetic algorithm first uses random search to find an initial population of feasible pumping schedules and then improve the pumping schedules through evolution mechanisms like crossover and mutation. Details about the genetic algorithm, including a discussion of its consistency in performance (Fig. S1), are provided in Supplementary Note 2.

**Hydraulic simulation**
EPANET uses a defined pumping schedule to simulate hydraulics in a WDN. The hydraulic simulator applies an iterative backward Euler method to calculate the flow rate of each link (e.g., pump, valve, pipe), the total head (i.e., the sum of elevation and pressure) of each node, and the operating status (i.e., flow, head gain, mechanical efficiency) of each pump. The simulated values are then input to the backtracking algorithm described below.

**Flow backtracking**
The backtracking algorithm consists of two steps – the first step is to backtrack water received by each consumer to its sources (Eq. 1-4); the second step is to backtrack the energy consumption or dissipation along the flow paths backtracked in the first step (Eq. 5-6). In Eq. 1, $i, j, t$ are indices of injection points, nodes in a WDN, and time steps, respectively. The MEI components associated with the transmission, treatment, and distribution of water are labeled as $\text{MEI}_i^{\text{Trans}}$, $\text{MEI}_i^{\text{Treat}}$, and $\text{MEI}_{i-j,t}^{\text{Dist}}$, which sum to $\text{MEI}_{j,t}$, the consumer node- and time-specific MEI value. Since we assume $\text{MEI}_i^{\text{Trans}}$ and $\text{MEI}_i^{\text{Treat}}$ to be constants for each injection point throughout the simulated time horizon, we can represent their sum as a constant $\text{MEI}_i^{\text{Pre-inj}}$ (Eq. 2). Besides the disaggregation of energy by stage of water supply, Eq. 1 also demonstrates the disaggregation of energy by source of water – $r_{i-j,t}$ is the fraction of water received by consumer $j$ during time step $t$ that comes from injection point $i$.

$$\text{MEI}_{j,t} = \sum_i r_{i-j,t} \cdot (\text{MEI}_i^{\text{Trans}} + \text{MEI}_i^{\text{Treat}} + \text{MEI}_{i-j,t}^{\text{Dist}}) \tag{1}$$

$$\text{MEI}_i^{\text{Pre-inj}} = \text{MEI}_i^{\text{Trans}} + \text{MEI}_i^{\text{Treat}} \tag{2}$$

Since we assume each node in a WDN to be a perfect mixer of upstream water, $r_{i-j,t}$ can be calculated from the values of $r_{i-k,t}$, where $k$ is the index of node $j$'s immediate upstream nodes (Eq. 3). In Eq. 3 and 4, $K_j$ is the set of immediate upstream nodes of node $j$, $Q_{k-j,t}$ is the flow rate from node $k$ to node $j$, and $Q_{j,t}$ is the total inflow rate at node $j$. It is worth noting that $Q_{j,t}$ is not necessarily the water consumption rate at node $j$. To solve the system of equations represented by Eq. 3, we use the injection points to create boundary conditions (i.e., $r_{i-i,t} = 1$ if injection point $i$ is actively injecting water).

$$k \in K_j \qquad r_{i-j,t} = \sum_k r_{i-k,t} \cdot \frac{Q_{k-j,t}}{Q_{j,t}} \tag{3}$$

$$Q_{j,t} = \sum_k Q_{k-j,t} \tag{4}$$

Once the values of $r_{i-j,t}$ are calculated, the next step is to calculate $\text{MEI}_{i-j,t}^{\text{Dist}}$, whose value is time-varying. If we define the absolute value of the head difference across nodes $k$ and $j$ as $h_{k-j,t}$, which can be directly retrieved from the simulation results, we can then solve for $\text{MEI}_{i-j,t}^{\text{Dist}}$ with Eq. 5 and 6. In Eq. 5, $H_{i-j,t}$ is sum of $h_{k-j,t}$ values along the path(s) from injection point $i$ to node $j$. It is worth noting that $h_{k-j,t}$ for a pump equals to the product of the actual head gain and the inverse of the mechanical efficiency. Similar to $r_{i-i,t}$, the calculation of $H_{i-j,t}$ propagates from upstream nodes to downstream nodes and requires boundary conditions at injection point (i.e., $H_{i-i,t} = 0$).

$$\text{MEI}_{i-j,t}^{\text{Dist}} = \rho \cdot g \cdot H_{i-j,t} \tag{5}$$

$$k \in K_j \qquad H_{i-j,t} = \frac{1}{\sum_k r_{i-k,t} \cdot Q_{k-j,t}} \cdot \sum_k r_{i-k,t} \cdot Q_{k-j,t} \cdot (H_{i-k,t} + h_{k-j,t}) \tag{6}$$

Since every node has its $H_{i-j,t}$ value, MEI can be calculated for all nodes, including those that are not consumers (e.g., zero-demand junctions, tanks). For such non-consumer nodes, their MEI values, if calculated using Eq. 1-6, can be interpreted as the energy intensity that would be incurred if a consumer connected to the node withdraws water. For a tank that is receiving water and saving the water for later injection, the non-consumer MEI is simply its instantaneous pre-injection energy intensity.

**Pre-injection energy intensity of a tank**
Unlike energy storage, which contributes only fractionally to energy supply and can be modeled as a price-taker in the electricity market (*32*), tanks (i.e., water storage) in a WDN can be the primary sources of water supply during hours when some or all pumps are off. As a result, the proposed computational framework treats each tank that is discharging water as an injection point. There are two challenges in calculating the pre-injection energy intensity of a tank. The first is the temporal decoupling of the charging and discharging of tanks – the consumption of energy to fill up a tank takes place before the tank functions as an injection point. Second, assuming each tank to be a perfect mixer, this calculation requires backtracking the entire volume of water stored in the tank. Since water may be pumped into a tank before the simulated time horizon (unless all tanks are empty at $t = 0$), it is not feasible to backtrack the entire volume of stored water without assuming an arbitrary initial condition. To address these issues, we only backtrack the flows into a tank during the simulated time horizon and use the weighted average MEI (non-consumer MEI) of such inflows as the proxy for the true pre-injection energy intensity. In Eq. 7, $T$ is the set of tanks and $R$ is the set of reservoirs of treated water (i.e., non-tank injection points). It is worth noting that $\text{MEI}_i^{\text{Pre-inj}}$ in Eq. 7 is not necessarily associated with a non-tank injection point because a tank may receive water from other tanks. In Eq. 8, $Q_{n,t}$ is the inflow rate at tank $n$ during time step $t$ and $Q_n$ is the total inflow in the entire simulated time horizon.

$$\begin{array}{c} n \in T \\ i \in (T \cup R) \setminus n \end{array} \qquad \text{MEI}_n^{\text{Pre-inj}} = \sum_t \frac{Q_{n,t}}{Q_n} \sum_i r_{i-n,t} \cdot (\text{MEI}_i^{\text{Pre-inj}} + \text{MEI}_{i-n,t}^{\text{Dist}}) \tag{7}$$

$$Q_n = \sum_t max\{Q_{n,t}, 0\} \tag{8}$$

This approximation approach eliminates the need to assume an arbitrary initial condition for calculating the pre-injection energy intensity of each tank. In addition, if the net water supply throughout the simulated time horizon is 0 for each tank, the computed MEI values, when multiplied with the consumption rates of water, will sum up to the total energy consumption and dissipation in the WSS in the simulated time horizon.

**Acknowledgments:** We thank Dr. Peter Fiske for helpful feedback on this manuscript. **Funding:** The conceptual and computational work was supported by the start-up funds of M.S. Mauter and the dissertation research time of Y. Liu; portions of Prof. Mauter's time for manuscript preparation was supported by the National Alliance for Water Innovation (NAWI), funded by the U.S. Department of Energy, Office of Energy Efficiency and Renewable Energy (EERE), Advanced Manufacturing Office, under Funding Opportunity Announcement Number DE-FOA-0001905 through a subcontract to Stanford University; **Author contributions:** Liu and Mauter are co-authors and developers; **Competing interests:** Authors have filed a provisional patent on this work; **Data and materials availability:** All data is available in the main text or the supplementary information.

# Supplementary Materials

## Supplementary Note 1. Comparison to previous methods that compute high-resolution energy intensity of WSS

Compared to the method proposed in (*11*), which estimated the energy intensity of water supply at the resolution of individual pressure zones, our computational framework improves the resolution to individual consumers and features a rigorous modeling of tanks, which is absent in (*11*). Similar to batteries in a power supply system, tanks in a WSS can temporally decouple supply and demand, which complicates the backtracking of water flows and the computation of flow path-specific energy intensity of delivered water. In order to more accurately account for the embedded energy of water discharged from tanks to consumers, (*12*) treats energy as a conservative property (e.g., concentration of NaCl) and uses EPANET's function that simulates the movement of chemicals to obtain the 'concentration of energy' (energy intensity of water supply) for each consumer. Despite the high resolution of the results achieved by this EPANET-based method, treating energy as a conservative property requires the user of the method to specify a constant of a timeseries of pre-determined energy intensity of each pump before running the simulation, which ignores the fact that the energy intensity of a pump varies with its operating status (e.g., flow rate, efficiency). Assuming the inherently time-varying operating status of a pump to be static or known can result in a large distortion of the results. In contrast, MEI calculated by our computational framework reflects the instantaneous operating status of each pump and incorporates the frictional and minor losses, which are infeasible to include using the method of (*12*).

## Supplementary Note 2. Genetic algorithm

To solve for the energy cost-minimizing pumping schedules, we implement a genetic algorithm (GA) that is adapted from methods described in (*15–20*). GA is a widely used evolutionary algorithm that incrementally improves the solution through two mechanisms – crossover and mutation. Crossover takes fractions (i.e., values of a subset of decision variables) from two or more different solutions and merge the fractions into a new solution. We call a new solution generated by crossover an offspring. To avoid premature convergence to a sub-optimal solution, GA usually uses mutation to randomly modify newly generated offspring. To make sure that the best solution does not get worse from generation to generation, many GAs also include an elite-picking mechanism that directly carries the best solutions in each generation over to the next generation. Unlike mathematical programming-based methods, GA cannot rigorously enforce all constraints. As a result, most GA implementations, including ours, account for the constraints as penalty functions and add the values of such penalty functions to the objective function, which gives the fitness function. In other words, fitter solutions are those whose fitness function values are lower, which suggest that both their objective function and penalty function values tend to be low. Details about our implementation of GA in this study are discussed in the following sub-sections.

### Fitness function

The fitness function value $F$ is the sum of four components – normalized electricity cost ($C_{elec}$), penalty for low tank levels ($P_{tank}$), penalty for detection of under-pressurized water delivery ($P_{pressure}$), and penalty for large deviation from the target fraction of water supply by each source ($P_{fraction}$). Given a timeseries of electricity prices, we first compute the total electricity cost by multiplying the hourly electricity consumption $E_t$ by the hourly electricity price $c_t$. To avoid poor scaling under different electricity prices, we normalize the total electricity cost by dividing it by the daily average electricity price. Subscript $t$ in Eq. S2 represents time. Similarly, the penalty function for low tank levels at the end of the day are also normalized. In the numerator of Eq. S3, $H_{T,f}$ and $H_{T,i}$ are final and initial water level of tank $T$, respectively. In the denominator, $H_{T,max}$ and $H_{T,min}$ bound the entire range of tank level fluctuation. Eq. S4 is the penalty function that penalizes delivered water with pressure lower than the required minimum pressure $p_{min}$ (i.e., 20 psi/14.06 m). $p_{low}$ is the lowest pressure of delivered water across all consumers throughout the day. In Eq. S5, $f_{target,i}$ and $f_{actual,i}$ are the target and actual fraction of water injection into the WDN through injection point $i$. $I_{Threshold,i}$ is an indicator function whose value is 1 when the difference between $f_{target,i}$ and $f_{actual,i}$ are greater than 0.02 and assumes a value of 0 otherwise. For example, if we plan to let injection point 1 account for 30% of the daily water injection (i.e., $f_{target,1} = 0.3$) but our pumping schedule results in a $f_{actual,1}$ value of 0.35, then the indicator function is activated. In Eq. S2-S5, $E_t$, $H_{T,f}$, $p_{low}$ and $f_{actual,i}$ are not constants with known values but can be retrieved from the results of the hydraulic simulation of a given pumping schedule.

$$F = C_{elec} + P_{tank} + P_{pressure} + P_{fraction} \tag{S1}$$

$$\forall t \qquad C_{elec} = \frac{24}{\sum_t c_t} \cdot \sum_t E_t \cdot c_t \tag{S2}$$

$$\forall T \qquad P_{tank} = \sum_T ((\frac{H_{T,f} - H_{T,i}}{H_{T,max} - H_{T,min}} + 0.2) \cdot 100)^2 \tag{S3}$$

$$P_{pressure} = (p_{min} - p_{low})^2 \cdot 10 \tag{S4}$$

$$P_{fraction} = \sum_i (f_{actual,i} - f_{target,i})^2 \cdot I_{Threshold,i} \cdot 250000 \tag{S5}$$

**Generate the initial population**
Our GA has 500 generations and each generation consists of 500 feasible solutions (i.e., pumping schedules). Since all pumps are fixed-speed in our case study, their status can be represented as 0-1 integers – 0 indicates 'shut down' and 1 indicates 'operating'. As a result, each solution consists of 11 strings of 0-1 integers and each string has 24 integers. In other words, each solution can be represented as a 11 by 24 matrix consisted of 0s and 1s. To generate the initial population, we randomly and repeatedly generate such 0-1 matrices until we obtain 500 feasible pumping schedules, which result in successful hydraulic simulations by EPANET.

**Crossover**
After the first generation, each new generation goes through a crossover stage. Out of the 500 new solutions, five are directly carried over from the best five solutions in the previous generation, the remaining 495 new solutions are all generated through crossover. After ranking the 500 solutions in the previous generation by fitness function value, we randomly select pairs of 'parent solutions' from the best 250 solutions (parent pool) and have each pair generate an offspring through a two-point crossover (*19*). For each crossover, the crossover points are randomly chosen but are the same for both parent solutions. We implement a roulette wheel selection here – for solution $j$ in the parent pool, its probability of being selected as a parent solution is set to be proportional to $(500 - rank_j)$, where $rank_j$ is the ranking of solution $j$ (*33*). If the crossover of a pair of randomly selected parent solutions generates an infeasible pumping schedule that leads to an unsuccessful hydraulic simulation, we re-select a pair of parents until a feasible new pumping schedule is obtained. It is worth noting that each solution can be selected as a parent multiple times and there is no guarantee that every solution in the parent pool will be selected at once.

**Mutation**
Once a feasible new pumping schedule is generated through crossover, there is an 80% probability that the pumping schedule is selected to undergo a mutation. For each mutation, 1-4 pumps are randomly chosen first. Afterwards, pump status over 1-6 randomly chosen and adjacent time steps (i.e., hours) are changed to either 0s or 1s for each pump selected to undergo mutation. The number of pumps to choose, the number of time steps to choose, and the pump status to change to are all random variables with uniform probability distribution.

**Parallel computing**
To accelerate the GA, we parallelize the crossover and mutation mechanisms. Instead of generating 495 new solutions one by one, we launch separate and parallel processes that each works towards finding a new feasible solution through crossover and mutation. After this parallelization, hydraulic simulations of different pumping schedules, which are the most time-consuming computational task in the GA, can be completed by parallel CPUs.

**Consistency of the algorithm**
As a heuristic method, GA cannot solve a problem to a given optimality gap. Instead, GA converges to a near-optimal solution efficiently. Therefore, it is important to check the reliability of the implemented GA and make sure that it produces consistent solutions given the same optimization inputs. Such consistence is particularly critical as we are using the GA to compute energy cost-minimizing pumping schedules for our sensitivity analysis. To test the consistency of the GA, we run the algorithm five times for the base case and compare the results. As Fig. S1 shows, the five optimized pumping schedules are very similar in their pattern and they result in very close daily energy costs, which are $56.8, $57.6, $57.8, $57.7, $57.8.

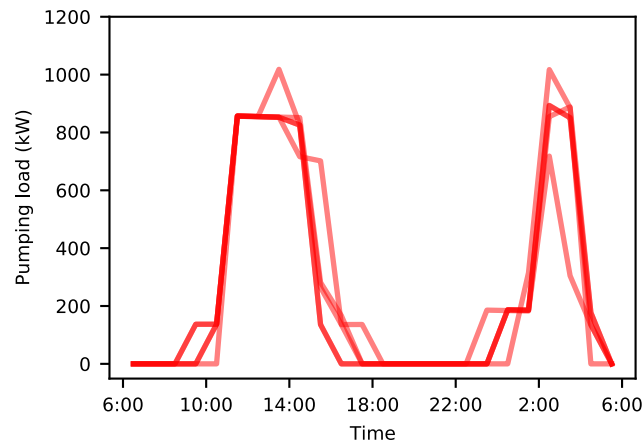

**Fig. S1**. Comparison of five pumping schedules optimized by the implemented genetic algorithm under the base case scenario.

## Supplementary Note 3. Details of the three-source water distribution network

The three-source WDN shown in Fig. 2A is adapted from a real-world system in Kentucky. The real-world WDN has 287,609 ft (87.7 km) of pipe and provides 1.04 million gallons (4581 m$^3$) of water per day to 2682 consumers. The skeletonized network maintains the overall configuration of the WDN but merges some nodes and pipes, which results in 331 consumers (one consumer is converted to a non-consumer junction as shown in Table S1).

As described in the main manuscript, we manipulate the locations of injection points I1, I2, and I3 from their true locations to improve their visualization in Fig. 2A. The true locations of the three injection points relative to the rest of the WDN are shown in Fig. S2. As Fig. S2 shows, I1 is closer to its downstream consumers than shown in Fig. 2A. On the other end of the WDN, I3 is much farther away from its downstream consumers than shown in Fig. 2A. For all the calculations in this study, we use the true locations of the injection points as well as their downstream pumps, pipes, and valves.

We make several small modifications to the nomenclature of the original network. The original WDN is downloaded as an EPANET file (*22*). In the original WDN, there are four reservoirs named R-1, R-2, R-3, and WTP. All reservoirs have known water levels and elevations. From the given information in the original EPANET file, it is obvious that R-2, R-3 are reservoirs storing water that requires no extra treatment for potable use, and WTP treats and stores water from its upstream reservoir R-1. We remove all the hydraulic components upstream WTP, including R-1, and rename WTP, R2, and R3 as I1, I2, and I3, respectively. Pumps originally named Pump 1, Pump 2, Pump 3, and Pump 4 are renamed as P2, P3, P1, and P4, respectively.

In addition to the removal of R-1 and the renaming of components, five additional modifications to the EPANET file are made for the base case scenario (shown in Table S1). In the original EPANET file, only a rated power is provided for each pump and all pumps are assumed to operate at a constant mechanical efficiency of 75%, which is unrealistic. To mimic more realistic operations of pumps, we use a previously proposed method (*34*) to re-construct reasonable pump curves for the pumps. Such curves include two sets – one set is head-flow ($H - Q$) curves and the other set is efficiency-flow ($\eta - Q$) curves. According to (*34*), the two curves for each pump can be estimated from the best-efficiency point (BEP), which is a pair of operating head and flow rate values. We run 1000 feasible one-time-step hydraulic simulations for the WDN under randomized conditions (e.g., initial tank levels, water demand) and use the mean head and flow rate values as its BEP. In the original EPANET file, P2 only has a rated power of 3 horsepower, which makes it impossible to withdraw large amounts of water from I2. We change the rated power of P2 to 150 horsepower before running the 1000 simulations. The re-constructed pump curves are shown in Fig. S3 below. Since the re-construction of $H - Q$ and $\eta - Q$ curves use the same two functions whose independent variables are both the ratio of flow rate to the BEP, the curves shown in Fig. S3 share the same shape but not the same range of values. Also, it is worth noting that pumps P3 and P4, which are connected in series, have very similar flow rates at their respective BEPs.

**Table S1.** Modifications to the EPANET file of the three-source WDN

| Modification 1 | Increase the water demand of all consumers at all times by a factor of four |
|---|---|
| Modification 1 | Make junction J-106 a non-consumer junction |

| Modification 2 | Move the start time of the time horizon (for hydraulic simulation) from 12 am to 6 am |
| Modification 3 | Increase the elevation of I1, I2, I3 and tanks T1, T2, T3 by 30 ft (9.14 m) |
| Modification 4 | Re-construct head-flow and efficiency-flow curves for pumps P1, P2, P3, and P4 |

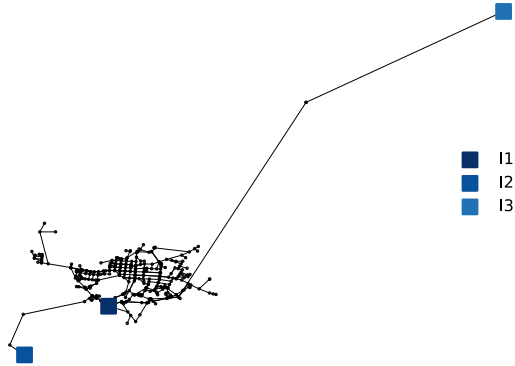

**Fig. S2. Locations of injection points I1, I2, and I3 relative to the rest of the WDN.** In the case study, all head losses in pipes are computed using the true pipe lengths that are proportional to the black lines in this figure.

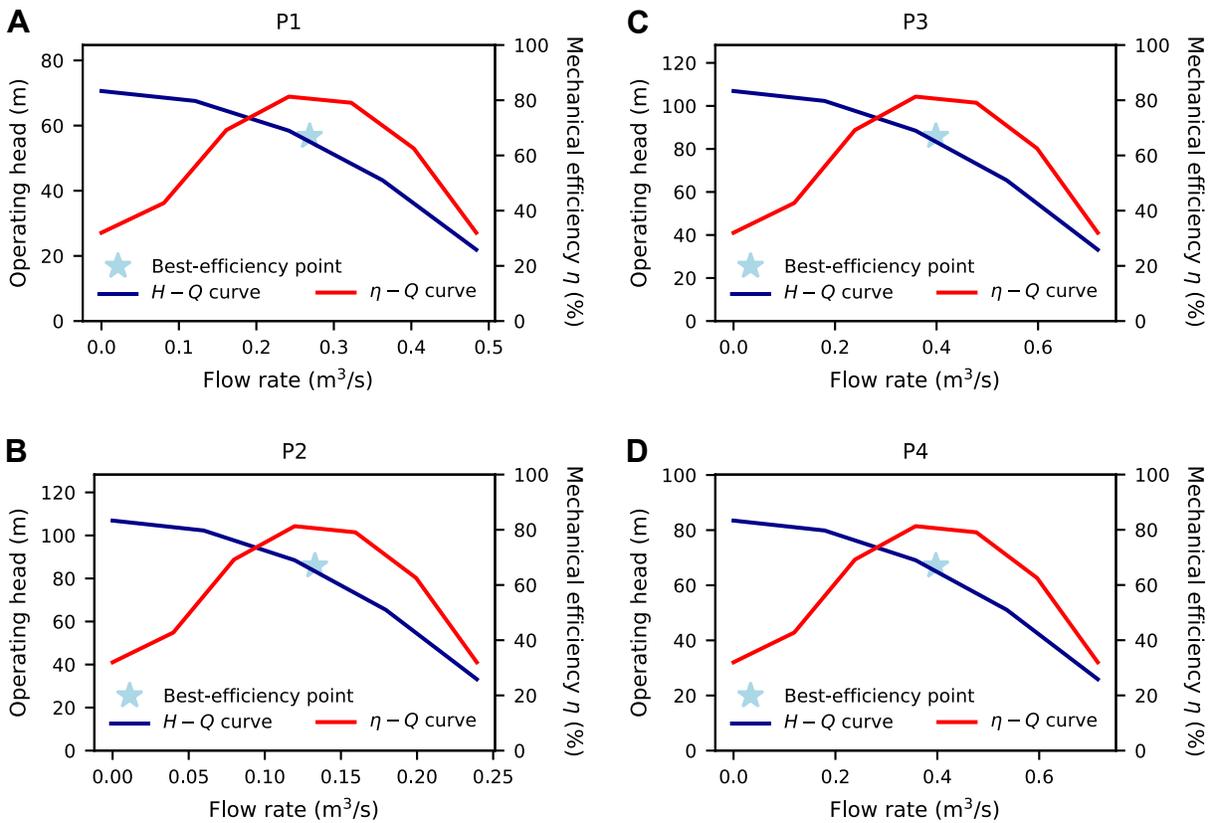

**Fig. S3. Re-constructed pump curves.** Head-flow curve and efficiency-flow curve for each pump in the three-source WDN. The curves are re-constructed from the best-efficiency points, which are obtained through hydraulic simulations with randomized conditions.

# Supplementary Note 4. Electricity prices used in the case study

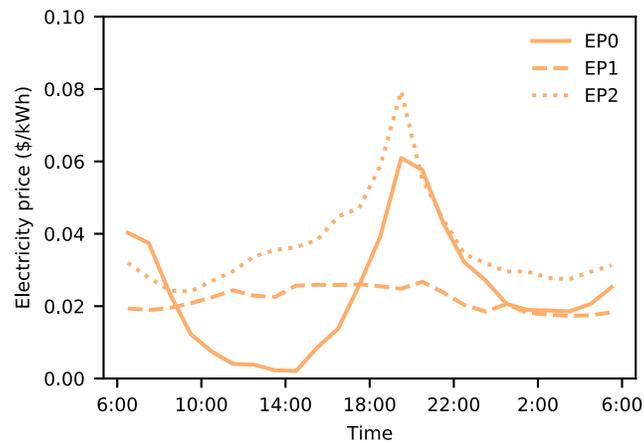

**Fig. S4. Electricity prices used in the case study.** EP0, EP1, and EP2 are averaged to hourly resolution from 15-minute-resolution data downloaded from (*35*), (*36*), and (*37*), respectively. EP0 is a 'duck curve' in which the electricity price reaches the lowest point in the afternoon when the power generation from solar farms are at its peak. EP1 is a relatively flat price timeseries. EP2 is a single-peak electricity price timeseries in which the electricity price peaks around evening.

# Supplementary Note 5. Hourly MEI values in the sensitivity analysis

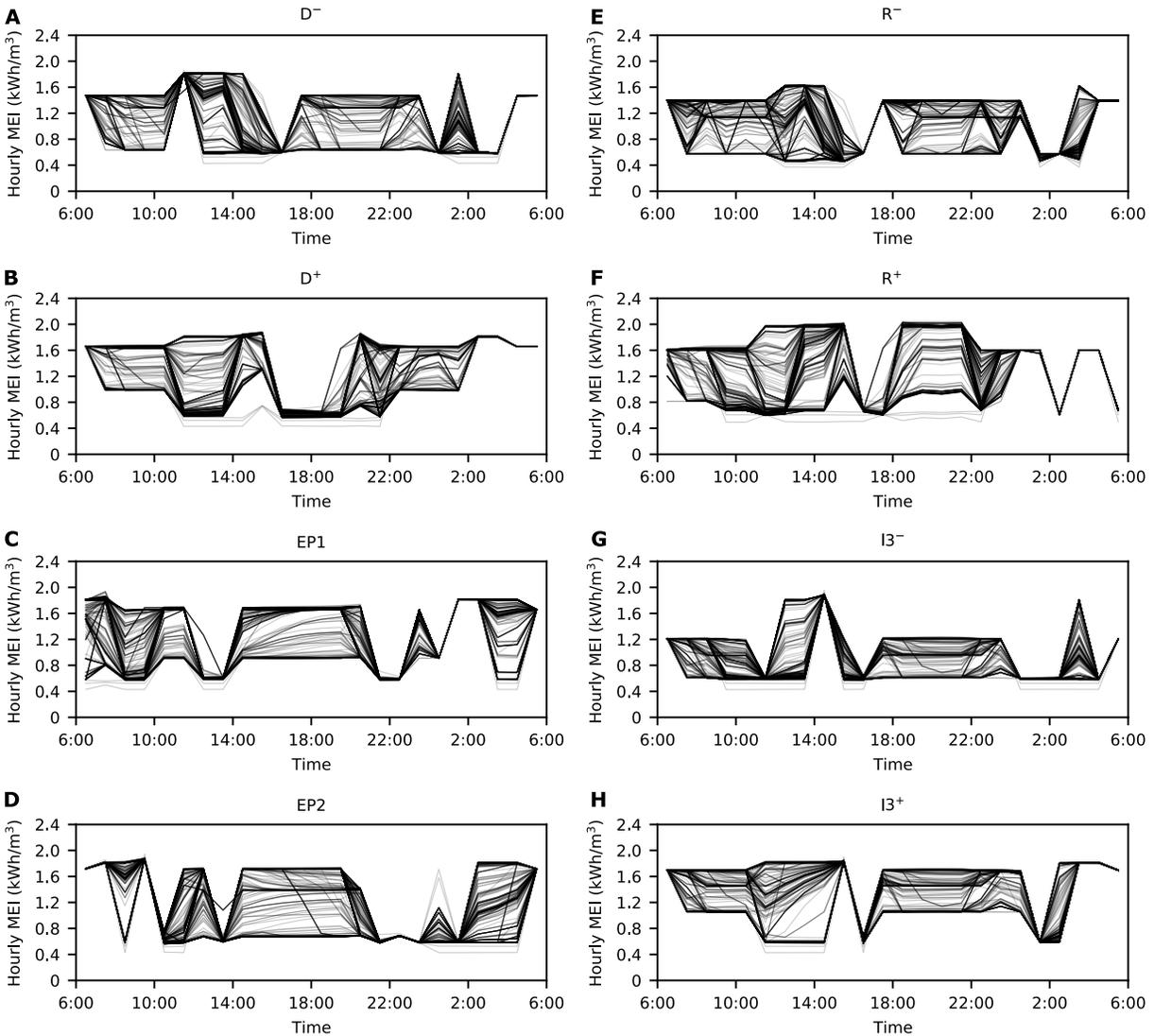

**Fig. S5. Hourly MEI of the 330 consumers in the three-source WDN in the eight scenarios of the sensitivity analysis**. 'D$^-$' and 'D$^+$' are scenarios where the hourly water consumption rates (i.e., demand) of all consumer nodes are decreased or increased by 25%, respectively. 'EP1' and 'EP2' are scenarios where alternative electricity price profiles are used to optimize the pumping schedule. 'R$^-$' and 'R$^+$' are scenarios where the roughness of all pipes is decreased or increased by 50%, respectively. 'I3$^-$' and 'I3$^+$' are scenarios where the target percentage of daily water injection through I3 is adjusted from 50% in the base case to 30% or 70%, respectively.

## Supplementary Note 6. Perturbation analysis of MEI

Since the water consumption rate of a single consumer is marginal (i.e., negligible) compared to the total injection rate through an injection point or flow rate in a water main, MEI is inherently insensitive to perturbations in the water consumption behavior of a single or a small number of consumer nodes. Such insensitiveness, or marginality of MEI, is demonstrated in Fig. S6. In this perturbation analysis, we randomly pick five consumers from the three-source WDN and vary their hourly water consumption rate by 10%, 20%, and 40% respectively. At each level of perturbation, the consumption rate of each perturbed consumer node could either increase or decrease, with equal probabilities, by the specified percentage. We repeat the perturbation at each level ten times, which generates 50 lines per level in Fig. S6. Since such perturbations have negligible effect on the overall water demand in the WDN, the base case pumping schedule remains feasible and is used to simulate water flows and calculate MEI values.

As the deviations in hourly MEI values suggest, even if the five consumer nodes' water consumption rates are perturbed by 40%, their hourly MEI values remain virtually unchanged compared to the unperturbed condition (Fig. 3c). The maximum deviation from the base case MEI values among the 150 lines in Fig. S6 is merely 0.077 kWh/m$^3$, which corresponds to a 5.7% increase. However, it is worth noting that larger perturbation in the water consumption behavior does lead to larger deviation in MEI.

Potential users of MEI and the computational framework must pay close attention to the marginality of MEI. On one hand, for example, the small deviations shown in Fig. S6 prove that the energy saving potential of a water-saving toilet can be directly computed by multiplying the water demand per flush by the MEI at the location. On the other hand, one should be careful when doing such multiplication over a large number of water consumers whose water consumption behaviors significantly deviate from the base case scenario where the MEI values are calculated. Once the change in water supply or demand is no longer marginal, the MEI values calculated before the change become inaccurate and need to be re-calculated.

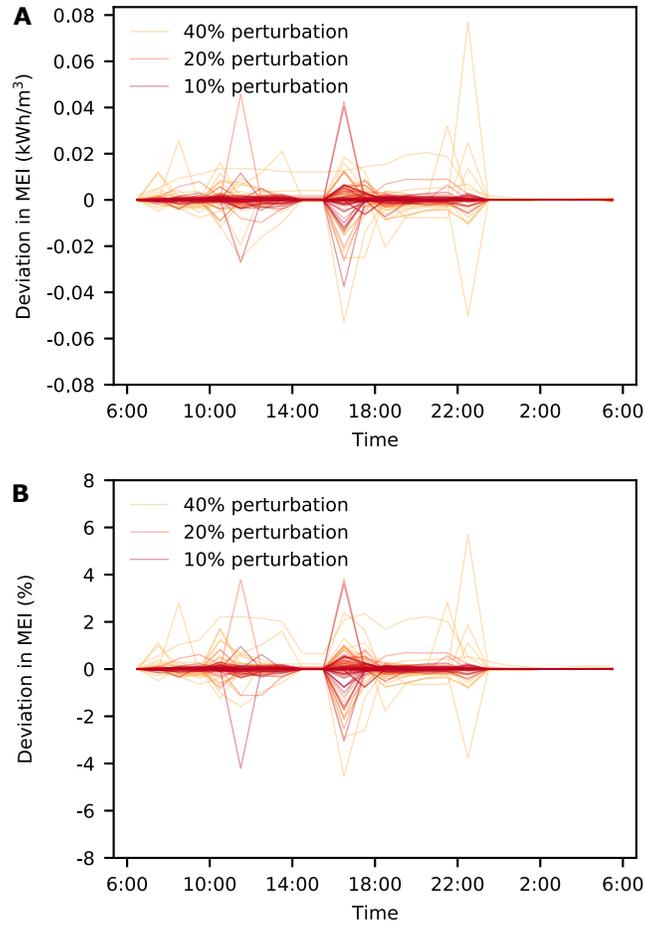

**Fig. S6. Results of the perturbation analysis.** Hourly MEI values of perturbed consumer nodes whose hourly water consumption rates are perturbed by 10%, 20%, and 40%. **A,** Deviation in hourly MEI values after perturbation measured in kWh/m$^3$. **B,** Deviation in hourly MEI values after perturbation measured in %.

## Supplementary Note 7. Hazen-Williams formula

The Hazen-Williams formula is a classical formula that correlates the head loss with the length, diameter, roughness of a pipe and the flow rate in the pipe. In Eq. S6, $h$ is the head loss (m), $L$ is the pipe length (m), $Q$ is the flow rate (m³/s), and $d$ is the pipe diameter (m). $C$ is the roughness coefficient, which is inversely proportional to the roughness of a pipe.

$$h = \frac{10.67 \cdot L \cdot Q^{1.852}}{C^{1.852} \cdot d^{4.87}} \tag{S6}$$

As Eq. S6 suggests, longer, thinner, and rougher pipes result in larger head losses. Therefore, by increasing the pipe roughness in the sensitivity analysis, we essentially achieve the same effect as stretching the pipes and expanding the size of the WDN. In the sensitivity analysis in the main manuscript, as we increase the roughness of each pipe by 50% (i.e., roughness coefficient decreases by 33.3%), the head loss increases by 112% with all else held equal.

## Supplementary Note 8. Details of the c-town water distribution network

Similar to the three-source WDN, the c-town WDN (Fig. 4A) is also adapted from an EPANET file with several modifications, which are listed in Table S2. This WDN is widely used as a benchmark system for studies that aim to optimize the design and operation of water distribution systems.

The original EPANET file comes with head-flow curves for the pumps but assumes a constant mechanical efficiency. As a result, we follow the same procedure applied to the three-source WDN to re-construct the efficiency-flow curves. Due to the similarity in both the procedure and the shapes of the resulting curves, we do not visualize each of the re-constructed curves for all eleven pumps. Using the GA described above, we optimize the pumping schedule for this WDN and plot the hourly pumping load, hourly water injection through the injection point and tanks, and hourly MEI values in Fig. S7.

**Table S2.** Modifications to the EPANET file of the c-town WDN

| Modification 1 | Move the start time of the time horizon (for hydraulic simulation) from 12 am to 6 am |
|---|---|
| Modification 2 | Increase the elevation of the only injection point and the 7 tanks by 30 ft (9.14 m) |
| Modification 3 | Re-construct efficiency-flow curves for the eleven pumps |

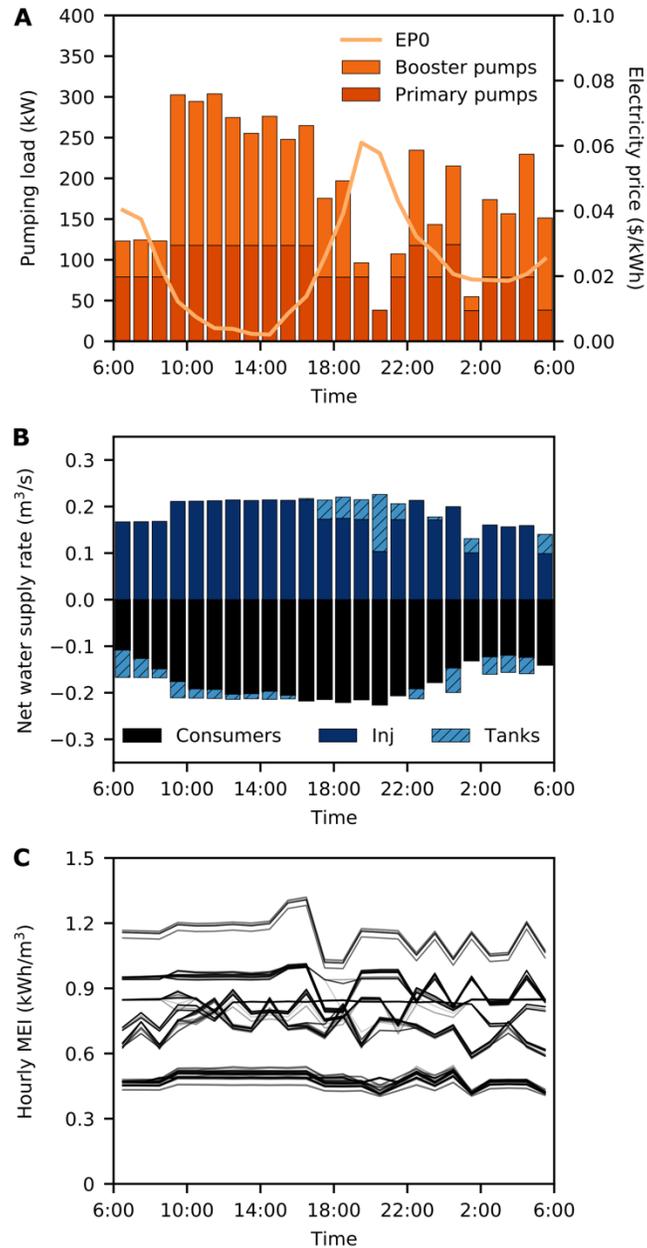

**Fig. S7. Pumping schedule and hourly MEI values of the c-town WDN. A.** Hourly load profile of the optimal pumping schedule that minimizes the electricity cost under electricity price profile EP0. Primary pumps are the three parallel pumps that are immediately downstream the injection point. Booster pumps are the remaining eight pumps. **B.** Hourly water injection through the injection point and the net water flows into tanks and consumers. 'Inj' is the abbreviation for 'injection point'. **C.** Hourly MEI values of 334 consumer nodes.